\documentclass[aps,prb,twocolumn,showpacs,floatfix]{revtex4}
\usepackage{graphicx}
\usepackage{graphicx}
\usepackage{amsmath}
\usepackage{amssymb}

\newcommand{\psiv}{\mbox{\boldmath$\psi$}}
\newcommand{\phiv}{\mbox{\boldmath$\varphi$}}

\begin{document}

\title{Electromagnons and instabilities in magnetoelectric materials with non-collinear spin orders}
\author{M. A. van der Vegte, C. P. van der Vegte, and M. Mostovoy}
\affiliation {Zernike Institute for Advanced Materials, University of Groningen,
Nijenborgh 4, 9747 AG Groningen, The Netherlands}
\date{\today}


\begin{abstract}
We show that strong electromagnon peaks can be found in absorption spectra of non-collinear magnets exhibiting a linear magnetoelectric effect. The frequencies of these peaks coincide with the frequencies of antiferromagnetic resonances and the ratio of the spectral weights of the electromagnon and antiferromagnetic resonance is related to the ratio of the static magnetoelectric constant and magnetic susceptibility. Using a Kagom\'{e} lattice antiferromagnet as an example, we show that frustration of spin ordering gives rise to magnetoelastic instabilities at strong spin-lattice coupling, which transform a non-collinear magnetoelectric spin state into a collinear multiferroic state with a spontaneous electric polarization and magnetization. The Kagom\'{e} lattice antiferromagnet also shows a ferroelectric incommensurate-spiral phase, where polarization is induced by the exchange striction mechanism.
\end{abstract}

%75.80.+q Magnetomechanical and magnetoelectric effects, magnetostriction
%71.45.Gm   Exchange, correlation, dielectric and magnetic response functions, plasmons
%78.20.-e Optical properties of bulk materials and thin films
%76.50.+g Ferromagnetic, antiferromagnetic, and ferrimagnetic resonances;
%75.10.Hk Classical spin models
%75.30.Kz   Magnetic phase boundaries (including magnetic transitions, metamagnetism, etc.)

\maketitle

\section{Introduction}
\label{sec:introduction}

The recent renewal of interest in multiferroic materials led to discovery of many novel compounds where electric polarization is induced by ordered magnetic states with broken inversion symmetry.\cite{CheongNatMat2007,KimuraARMR2007,RameshNatMat2007}
The electric polarization in multiferroics is very susceptible to changes in spin ordering produced by an applied magnetic field, which gives rise to dramatic effects such as the magnetically-induced polarization flops and colossal magnetocapacitance.\cite{KimuraNature2003,HurNature2004,GotoPRL2004} Magnetoelectric interactions also couple spin waves to polar phonon modes and make possible to excite magnons by an oscillating electric field of light, which gives rise to the so-called electromagnon peaks in photoabsorption.\cite{SmolenskiiSPU1982}

Electromagnons were recently observed in two groups of multiferroic orthorombic manganites, $R$MnO$_3$ ($R$ = Gd,Tb,Dy,Eu$_{1-x}$Y$_{x}$)  and $R$Mn$_2$O$_5$ ($R$ = Y,Tb). \cite{PimenovNaturePhys2006,PimenovPRB2006,ValdesPRB2007,SushkovPRL2007}  Ferroelectricity in $R$MnO$_3$ appears in a non-collinear antiferromagnetic state with the cycloidal spiral ordering and the magnetoelectric coupling originates from the so-called inverse Dzyaloshinskii-Moriya  mechanism.  \cite{KatsuraPRL2005,KenzelmannPRL2005,SergienkoPRB2006,MostovoyPRL2006,MalashevichPRL2008} In Ref.~[\onlinecite{KatsuraPRL2007}] it was noted that the same mechanism can couple magnons to photons and that an oscillating electric field of light can excite rotations of the spiral plane.  However, the selection rule for the electromagnon polarization resulting from this coupling does not agree with recent experimental  data\cite{ValdesPRB2007,KidaPRB2008,SushkovJPCM2008,TakahashiPRL2008} and, moreover, the inverse Dzyaloshinskii-Moriya mechanism  of relativistic nature is too weak to explain the strength of the electromagnon peaks in $R$MnO$_3$.

These peaks seem to originate from the exchange striction, i.e. ionic shifts induced by changes in the Heisenberg exchange energy when spins order or oscillate.\cite{ValdesPRL2009} This mechanism explains the experimentally observed polarization of electromagnons. Since the Heisenberg exchange interaction is stronger than the Dzyaloshinskii-Moriya interaction, it can induce larger electric dipoles. In Ref.~[\onlinecite{ValdesPRL2009}] it was shown that the magnitude of the spectral weight of the giant electromagnon peak in the spiral state of rare earth manganites is in good agreement with the large spontaneous polarization in the E-type antiferromagnetic state,\cite{LorenzPRB2006} which has not been reliably measured yet but is expected to exceed the polarization in the spiral state by 1-2 orders of magnitude.\cite{SergienkoPRL2006,PicozziPRL99}

From the fact that the mechanism that couples magnons to light in rare earth manganites is different from the coupling that induces the static polarization in these materials we can conclude that electromagnons can also be observed in non-multiferroic magnets. In this paper we focus on electromagnons in materials exhibiting a linear magnetoelectric effect, i.e. when an applied magnetic field, $\mathbf{H}$, induces an electric polarization, $\mathbf{P}$, proportional to the field, while an applied electric field, $\mathbf{E}$, induces a magnetization, $\mathbf{M}$. This unusual coupling takes place in antiferromagnets where both time reversal and inversion symmetries are spontaneously broken.\cite{Landaubook1984,FiebigJAPD2005}

It is natural to expect that when an electric field applied to a magnetoelectric material oscillates, the induced magnetization will oscillate too. Such a dynamical magnetoelectric response, however, requires presence of excitations that are coupled both to electric and magnetic fields. They appear when magnons, which can be excited by an oscillating magnetic field (antiferromagnetic resonances), mix with polar phonons, which are coupled to an electric field. Thus in materials showing a linear magnetoelectric effect, for each electromagnon peak there is an antiferromagnetic resonance with the same frequency.

\begin{figure}
\includegraphics[scale=0.4]{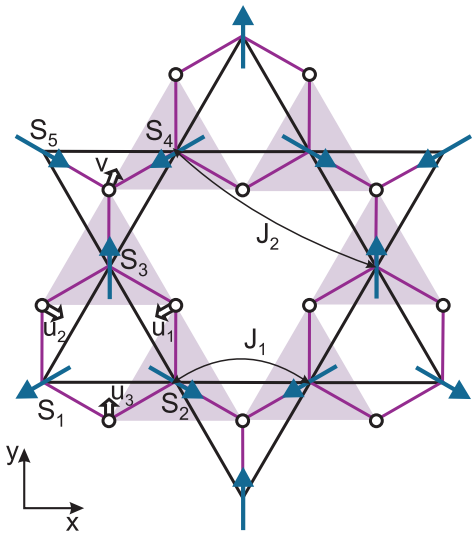}
\caption{(Color online) The Kagom\'{e} magnet with the KITPite crystal structure, in which the ligand ions (open circles) mediating the superexchange between spins are positioned in a way that gives rise to a strong linear magnetoelectric response in the $120^{\circ}$ spin state. Here,  $J_1$ and $J_2$ denote, respectively, the nearest-neighbor and next-nearest-neighbor exchange constants, the solid arrows denote spins, while the empty arrows denote the shifts of the ligand ions. }
\label{fig:model}
\end{figure}

This reasoning does not apply to all magnetoelectrics and the dc magnetoelectric effect is not necessarily related to hybrid spin-lattice excitations. As will be discussed below, in materials with collinear spin orders electromagnons either do not exist or have a relatively low spectral weight. In this paper we argue that electromagnons should be present in non-collinear antiferromagnets showing strong static magnetoelectric response.  As a simple example, we consider a Kagom\'{e} lattice antiferromagnet with the 120$^{\circ}$ spin ordering,  shown in Fig.~\ref{fig:model}. Such an ordering has a nonzero magnetic monopole moment, which allows for a linear magnetoelectric effect with the magnetoelectric tensor $\alpha_{ij} = \alpha \delta_{ij}$ for electric and magnetic fields applied in the plane of the Kagom\'{e} lattice.\cite{SpaldinJPCM2008}  A relatively strong magnetoelectric response was recently predicted for Kagom\'{e} magnets with the KITPite crystal structure, in which  magnetic ions are located inside oxygen bipyramids.\cite{DelaneyPRL2009} In this structure the oxygen ions mediating the superexchange in basal planes are located outside the up-triangles forming the Kagom\'{e} lattice and inside the down-triangles or vice versa (see Fig.~\ref{fig:model}), in which case magnetoelectric responses of all triangles add giving rise to a large magnetoelectric constant.

This paper is organized as follows. In Sec.~\ref{sec:symmetry} we analyze the symmetry of magnon modes and the magnetoelectric coupling in the Kagom\'{e} lattice magnet with the KITPite structure and show that the dc magnetoelectric effect in this system is related to presence of electromagnon modes. The common origin of the dc and ac magnetoelectric responses implies existence of relations between static and dynamic properties of magnetoelectric materials, derived in Sec.~\ref{sec:relations}.  In Sec.~\ref{sec:softening} we discuss softening of (electro)magnons and the resulting divergence of the coupled magnetoelectric response. In Sec.~\ref{sec:phasediagram} we discuss the transition from a magnetoelectric to a  multiferroic state at a strong spin-lattice coupling and plot the phase diagram of our model system. In section~\ref{sec:discussion} we discuss the importance of non-collinearity of spins for dynamic magnetoelectric response and possible electromagnons in known magnetoelectric materials. In Sec.~\ref{sec:conslusions} we conclude.

\section{Symmetry considerations}
\label{sec:symmetry}

\begin{figure}[htbp]
\includegraphics[scale=0.37]{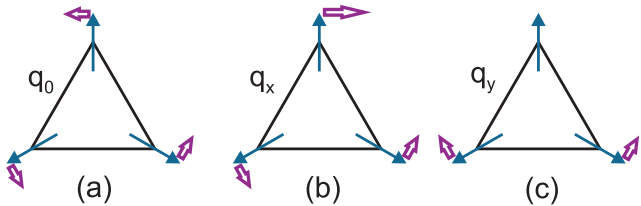}
\caption{(Color online) The magnon modes in a triangle with the $120^{\circ}$ spin ordering. The thin arrows indicate the directions of average spins, while the thick arrows show the in-plane components of the magnon modes.} \label{fig:modes}
\end{figure}

The coupling of magnetic excitations in the Kagom\'{e} magnet to electric field, resulting from the Heisenberg exchange striction or any other non-relativistic interaction, can be found using the method outlined in Ref.~[\onlinecite{ValdesPRL2009}]. To simplify notation,  we consider a single up-triangle, which has the same point symmetry as the whole Kagom\'{e} lattice with the $120^{\circ}$ spin ordering shown in Fig.~\ref{fig:model}. The form of the magnetoelectric coupling is constrained by the $3_z$ and $m_x$ symmetry operations:\cite{BulaevskiiPRB2008}
\begin{eqnarray}
H_{\rm me} &=& - \gamma
\left\{
\frac{E_{x}}{\sqrt{2}}
\left[
\left(\mathbf{S}_{2}\cdot\mathbf{S}_{3}\right) -
\left(\mathbf{S}_{1}\cdot\mathbf{S}_{3}\right)
\right]
\right.
\nonumber \\
&+&
\left.
\frac{E_{y}}{\sqrt{6}}
\left[
\left(\mathbf{S}_{1}\cdot\mathbf{S}_{3}\right) +
\left(\mathbf{S}_{2}\cdot\mathbf{S}_{3}\right)
- 2 \left(\mathbf{S}_{1}\cdot\mathbf{S}_{2}\right)
\right]
\right\}.
\label{eq:Hme1}
\end{eqnarray}
We then replace $\mathbf{S}_{i}$ by $\langle S \rangle \mathbf{n}_{i} +\delta \mathbf{S}_{i}$, where
the unit vectors  $\left(\mathbf{n}_{1},
\mathbf{n}_{2},\mathbf{n}_{3}\right) = \left(-\frac{\sqrt{3}}{2} {\hat x}-\frac{1}{2}{\hat y},\frac{\sqrt{3}}{2} {\hat x}-\frac{1}{2}{\hat y}, {\hat y}\right)$ describe the $120^{\circ}$ spin ordering in the $xy$ plane and   $\delta \mathbf{S}_{i} \perp \mathbf{n}_i$ is the oscillating part, which is the superposition of the orthogonal magnon modes in the triangle (the zero wave vector magnons for the Kagom\'{e} lattice):
\begin{equation}
\delta \mathbf{S}_{i} = \sum_{\alpha}\left(q_{\alpha} \psiv_{\alpha i} +  \langle S \rangle p_{\alpha} \phiv_{\alpha i}  \right),
\end{equation}
where $\alpha = 0,x,y$ labels the magnon,
\begin{equation}
\left\{
\begin{array}{rcl}
\phiv_{0i} &=& {\hat z}\frac{1}{\sqrt{3}}\left(1,1,1\right),\\
\phiv_{xi} &=& {\hat z} \frac{1}{\sqrt{6}}\left(1,1, -2\right),\\
\phiv_{yi} &=& {\hat z} \frac{1}{\sqrt{2}}\left(-1,1,0\right),
\end{array}
\right.
\end{equation}
are the out-of-plane components of the magnons and $\psiv_{\alpha i} = \phiv_{\alpha i} \times \mathbf{n}_i$ are the in-plane components (see Fig.~\ref{fig:modes}),
\begin{equation}
\left\{
\begin{array}{rcl}
\psiv_{0} &=& \frac{1}{\sqrt{3}}\left(\frac{1}{2} {\hat x}-\frac{\sqrt{3}}{2}{\hat y},\frac{1}{2} {\hat x}+\frac{\sqrt{3}}{2}{\hat y}, -{\hat x}\right),\\ \psiv_{x} &=& \frac{1}{\sqrt{6}}\left(\frac{1}{2} {\hat x}-\frac{\sqrt{3}}{2}{\hat y},\frac{1}{2} {\hat x}+\frac{\sqrt{3}}{2}{\hat y}, 2{\hat x}\right),\\
\psiv_{y} &=& \frac{1}{\sqrt{2}}\left(-\frac{1}{2} {\hat x}+\frac{\sqrt{3}}{2}{\hat y},\frac{1}{2} {\hat x}+\frac{\sqrt{3}}{2}{\hat y}, 0\right).
\end{array}
\right.
\end{equation}

The single-magnon excitation by an electric field is described by the terms linear in $\delta \mathbf{S}$, while the terms quadratic in $\delta \mathbf{S}$ give rise to the photoexcitation of a two-magnon continuum.  Since spins order in plane, the polarization oscillations are induced by the in-plane oscillations of spins and the coupling of electric field to magnons, obtained from Eq.(\ref{eq:Hme1}), has the form:
\begin{equation}
H_{\rm me} = - g_E  \left(q_x E_x + q_y E_y\right),
\label{eq:Hme2}
\end{equation}
where $g_E = \frac{3}{2}\gamma \langle S \rangle$. This magnetoelectric coupling is only nonzero in the magnetically ordered state with broken time reversal symmetry, which is why the coupling constant is proportional to $\langle S \rangle$.

We note that Eq.(\ref{eq:Hme2})  can also be obtained by a conventional symmetry analysis of magnetic modes. The vector $\psiv_0$ and, hence, the corresponding amplitude, $q_0$, forms a one-dimensional representation $\Gamma_1$, while  $\left(\psiv_{x},\psiv_{y}\right)$ and $\left(\!
\begin{array}{c}
q_{x}\\
q_{y}
\end{array}
\!\right)$ form a two-dimensional representation $\Gamma_3$ (see Table \ref{tab:ireps}). The direct product, $\Gamma_2 \times \Gamma_3$, where $\Gamma_2$ is the symmetry of the spin ordering, transforms as the doublet of the in-plane components of the electric fields, $\Gamma_4$, which leads to Eq.(\ref{eq:Hme2}). This general symmetry analysis is, however, insensitive to the microscopic mechanism of the  magnetoelectric coupling, whereas the derivation staring from Eq.(\ref{eq:Hme1}) only applies to non-relativistic mechanisms.

\begin{table}[htbp]
\centering
\begin{tabular}{|c|c|c|c|c|}
\hline
 & & $3_z$ & $m_x$ &  $T$\\
[1.0ex]
\hline
$\Gamma_1$&
$q_{0}$ & +1 & +1 & $-1$\\
[0.4ex]
$\Gamma_2$&$\left\langle\mathbf{S}\right\rangle$ & +1 & $-1$ & $-1$\\
[1.0ex]
$\Gamma_3$ &
$\left(\!
\begin{array}{c}
q_{x}\\
q_{y}
\end{array}
\!\right),
\left(\!
\begin{array}{c}
H_{x}\\
H_{y}
\end{array}
\!\right)$
& $\left(\!
\begin{array}{cc}
-\frac{1}{2} & - \frac{\sqrt{3}}{2}\\
+ \frac{\sqrt{3}}{2} & -\frac{1}{2}
\end{array}
\!\!\right)$ &
$\left(\!
\begin{array}{cc}
+1 & 0\\
0& -1
\end{array}
\!\!\right)$
&
$\left(\!\!
\begin{array}{cc}
-1 & 0\\
0 & -1
\end{array}
\!\!\right)$\\
[3.0ex]
$\Gamma_4$ &
$
\left(\!
\begin{array}{c}
E_{x}\\
E_{y}
\end{array}
\!\right)$
& $\left(\!
\begin{array}{cc}
-\frac{1}{2} & - \frac{\sqrt{3}}{2}\\
+ \frac{\sqrt{3}}{2} & -\frac{1}{2}
\end{array}
\!\right)$ &
$\left(\!\!
\begin{array}{cc}
-1 & 0\\
0& +1
\end{array}
\!\!\right)$
&
$\left(\!\!
\begin{array}{cc}
+1 & 0\\
0 & +1
\end{array}
\!\!\right)$\\
[3.0ex]
\hline
\end{tabular}
\caption{The transformation properties of several irreducible representations of the space group of the Kagom\'{e} lattice and time reversal operation $T$.}
\label{tab:ireps}
\end{table}

Table~\ref{tab:ireps} shows that the coupling to the in-plane components of magnetic field has the form,
\begin{equation}
H_{\rm m} = - g_{H} \left(q_{x} H_{x} + q_{y} H_{y}\right),
\label{eq:Hm}
\end{equation}
while the Hamiltonian describing magnon modes with zero wave vector in the Kagom\'{e} layer is,
\begin{equation}
H(p,q) =  \frac{p_0^2}{2 m_0}  + \frac{1}{2m}\left(p_x^2 + p_y^2\right) + \frac{\kappa_0 q_0^2}{2}+\frac{\kappa}{2}\left(q_x^2 + q_y^2 \right).
\label{eq:Hspin}
\end{equation}
If we consider, for example, the microscopic spin Hamiltonian describing the antiferromagnetic nearest-neighbor and next-nearest-neighbor Heisenberg exchange interactions with the exchange constants, respectively, $J_1$ and $J_2$ and the easy plane magnetic anisotropy $\Delta$,
\begin{equation} \label{eq:Hamspin}
H = J_1 \sum_{\langle ij \rangle }  \mathbf{S}_i \cdot
\mathbf{S}_j +J_2 \sum_{\langle \langle ij \rangle \rangle} \mathbf{S}_i \cdot \mathbf{S}_j + \frac{\Delta}{2} \sum_i \left(S_{i}^{z}\right)^2,
\end{equation}
for which the $120^{\circ}$ spin ordering shown in Fig.~\ref{fig:model} is a classical ground state,\cite{HarrisPRB1992,ElhajalPRB2002} we get
\begin{equation}
\begin{array}{ll}
m_{0}^{-1} = \left[6(J + J') + \Delta\right]\langle S \rangle^2, & m^{-1} = \Delta \langle S \rangle^2,\\
\kappa_0 = 0, & \kappa = 3\left(J_1 + J_2\right).
\end{array}
\end{equation}

The linearized equations of motion for spins in applied electric and magnetic fields are obtained from Eqs.(\ref{eq:Hme2}),(\ref{eq:Hm}), and (\ref{eq:Hspin}), if we impose the commutation relations for the amplitudes of the in-plane and out-of-plane parts of ${\delta \mathbf{S}}$:
\begin{equation}
\left[q_{\alpha},p_{\beta}\right] = i \delta_{\alpha,\beta}.
\end{equation}
From these equations we find the frequencies of the three magnon modes with zero wave vector:  $\omega_0^2 = \kappa_0 m_0^{-1} = 0$ and $\omega_x^2 = \omega_y^2 = \kappa m^{-1} = 3\left(J_1 + J_2\right) \Delta \langle S \rangle^2$.

Minimizing the spin energy with respect to $q_x$ and $q_y$ in the static limit, we obtain an effective magnetoelectric coupling,
\begin{equation}
H_{\rm me} = - \alpha \left(H_x E_x + H_y E_y\right).
\label{eq:Fme}
\end{equation}
where the magnetoelectric coefficient $\alpha =  \frac{g_Eg_H}{2\kappa}$. Furthermore, the $q_{x}$-mode can be excited by both electric and magnetic field oscillating with the frequency $\omega_x$ in the direction parallel to the $x$ axis, while the $q_{y}$-mode can be excited by both $E_y$ and $H_y$, which shows that the static linear magnetoelectric effect in this non-collinear magnet is related to the presence of electromagnon and  antiferromagnetic resonance peaks with equal frequencies in the optical absorption spectrum.

\section{Relations between static and dynamic magnetoelectric response}
\label{sec:relations}

The common origin of the static and dynamic magnetoelectric response of non-collinear magnets leads to quantitative relations between dc susceptibilities and spectral weights of peaks in the optical absorption spectrum. These relations simplify when the coupling of magnons to electric field is mediated by a single optical phonon. The description of magnons in terms of conjugated coordinates and momenta is very convenient for derivation of these relations, since the coupled magnon and optical phonon are in this approach just a pair of coupled oscillators:
\begin{eqnarray}
H &=&  \frac{1}{2m}\left(p_x^2 + p_y^2\right) +\frac{\kappa}{2}\left(q_x^2 + q_y^2 \right) \nonumber \\
&~&+ \frac{1}{2M}\left(P_x^2 + P_y^2\right) +\frac{K}{2}\left(Q_x^2 + Q_y^2 \right) \nonumber \\
&~&-
\lambda \left(q_x Q_x + q_y Q_y \right) -
f \left(Q_x E_x + Q_y E_y \right) \nonumber \\
&~&- g_H \left(q_x H_x + q_y H_y \right),
\label{eq:Hcoupled}
\end{eqnarray}
where $(Q_x,P_x)$ and $(Q_y,P_y)$ are the coordinates and momenta of the optical phonons coupled to magnons, which also form a two-dimensional representation.

The magnetoelectric response of such a system is easy to calculate. The result can be expressed in terms of observable quantities, such as the `dressed' magnon and phonon frequencies, $\omega_{\rm mag}$ and $\omega_{\rm ph}$, and the spectral weights of the magnon and phonon peaks excited by an electric and magnetic field. We denote the spectral weight of the electromagnon peak by
\begin{equation}
S_{\rm mag}^{E} = 8 \int\!\!d\omega \omega {\chi}''_{\rm e}(\omega),
\label{eq:SmagE}
\end{equation}
where ${\chi}''_{\rm e}(\omega)$ is the imaginary part of the dielectric ac susceptibility, while the spectral weight of the antiferromagnetic resonance is,
\begin{equation}
S_{\rm mag}^{H} = 8 \int\!\!d\omega \omega {\chi}''_{\rm m}(\omega),
\label{eq:SmagH}
\end{equation}
where ${\chi}''_{\rm m}(\omega)$ is the imaginary part of the magnetic ac susceptibility. The integration in Eqs.(\ref{eq:SmagE}) and Eq.(\ref{eq:SmagH}) goes over an interval of frequencies around $\omega_{\rm mag}$. The two spectral weights for the phonon $S_{\rm ph}^{E}$ and $S_{\rm ph}^{H}$ are defined in a similar way. We assume that the magnon and phonon peaks are sufficiently narrow and can be separated from each other. The four spectral weights satisfy a relation,
\begin{equation}
S_{\rm mag}^{E} S_{\rm mag}^{H} = S_{\rm ph}^{E} S_{\rm ph}^{H},
\label{eq:weights}
\end{equation}
following from the fact that an electric field only interacts with the `bare' phonon, while a magnetic field is only coupled to the `bare' magnon.

The relations between the dc and ac magnetoelectric responses of the coupled spin-lattice system have the form,
\begin{equation}
\left\{
\begin{array}{rcl}
\Delta {\epsilon} &=&
%4\pi {\chi}'_{\rm e}(0) =
\frac{S_{\rm mag}^{E}}{\omega_{\rm mag}^2}+
\frac{S_{\rm ph}^{E}}{\omega_{\rm ph}^2},\\ \\
\Delta {\mu} &=&
%4\pi {\chi}'_{\rm m}(0) =
\frac{S_{\rm mag}^{H}}{\omega_{\rm mag}^2}+
\frac{S_{\rm ph}^{H}}{\omega_{\rm ph}^2},\\ \\
4 \pi \vert \alpha \vert &=& \sqrt{S_{\rm mag}^{E}S_{\rm mag}^{H}}
\left|\frac{1}{\omega_{\rm mag}^2}-
\frac{1}{\omega_{\rm ph}^2}\right|,
\end{array}
\right.
\label{eq:KK}
\end{equation}
where $\Delta {\epsilon}$($\Delta {\mu}$) is the increase of the
real part of the dielectric constant (magnetic permeability) at
zero frequency resulting from the magnon and phonon peaks (we use
the Gaussian units). The first two equations are the
Kramers-Kronig relations for the real and imaginary parts of
dielectric and magnetic susceptibilities, while the last relation
follows from equations of motion. Combining Eqs.(\ref{eq:weights})
and (\ref{eq:KK}) we can express the ratio of the spectral weights
of the electromagnon and the antiferromagnetic resonance through
the ratio of the static magnetoelectric constant $\alpha$ and
magnetic susceptibility ${\chi}_{\rm m} = {\chi}'_{\rm
m}(\omega=0)$:
\begin{equation}
\frac{S_{\rm mag}^{E}}{S_{\rm mag}^{H}} =
\left( \frac{\alpha}{{\chi}_{\rm m}} \right)^2
\left(\frac{1+\frac{\omega_{\rm mag}^2}{\omega_{\rm ph}^2} \frac{S_{\rm mag}^{E}}{S_{\rm ph}^{E}}}{1 - \frac{\omega_{\rm mag}^2}{\omega_{\rm ph}^2}}\right)^2
\label{eq:emagafmr}
\end{equation}
For  $\omega_{\rm mag}^2 \ll \omega_{\rm ph}^2$, the ratio of the spectral weights is just the square of the ratio of the dc magnetoelectric constant and magnetic susceptibility.

Due to the spin-lattice coupling, the `dressed' magnon    frequency, $\omega_{\rm mag}$, is lower than its `bare' value, $\sqrt{\frac{\kappa}{m}}$ (assuming that the `bare' magnon frequency is smaller than the `bare' phonon frequency). As the spin-lattice coupling increases, $\omega_{\rm mag}$ vanishes at a critical value of the coupling. According to Eq.(\ref{eq:KK}), this results in the simultaneous divergency of $\epsilon$, $\mu$ and $\alpha$, indicating an instability towards a multiferroic phase, which is both ferroelectric and ferromagnetic. Another manifestation of this instability is the fact that as the spin-lattice coupling approaches  the critical value, the magnetoelectric constant $\alpha$ tends to its upper bound equal the geometric mean of the dielectric and magnetic susceptibilities, $\sqrt{\chi_{\rm e} \chi_{\rm m}}$, imposed by the requirement of stability with respect to applied electric and magnetic fields.\cite{BrownPR1968}

\section{Magnon softening}
\label{sec:softening}

To study the transition from the magnetoelectric state of the Kagom\'{e} magnet to the multiferroic state in more detail, we consider a simple microscopic model, in which positions of magnetic ions are fixed, while ligand ions are allowed to move. The spin-lattice coupling originates from the dependence of the exchange constants on displacements of ligand ions mediating the superexchange in the direction perpendicular to the straight line connecting two neighboring spins. We denote the positions of the three ligand ions outside up-triangles  by $\mathbf{u}_1$,  $\mathbf{u}_2$, and  $\mathbf{u}_3$, while the position of a ligand ion inside a down-triangle is denoted by $\mathbf{v}$. Then, for example, the exchange constant for the spins $\mathbf{S}_{1}$ and $\mathbf{S}_{2}$ is
$
J_1+ J_{1}' \left(u_{3}\right)_{y},
$
while for the spins $\mathbf{S}_{4}$ and $\mathbf{S}_{5}$ it is
$
J_{1}+ J_{1}' v_{y}
$
(see Fig.~\ref{fig:model}). Furthermore, we assume that phonons are dispersionless and the lattice energy for a pair of the up- and down-triangles is,
\begin{equation}
U_{lat} = \frac{K}{2} \left(\sum_{i=1}^{3}
\mathbf{u}_{i}^2  +  \mathbf{v}^2\right),
\label{eq:Ulat}
\end{equation}
where $K$ is the spring constant.

For the $120^{\circ}$ structure shown in Fig.~\ref{fig:conf}(a), the magnetoelectric constant $\alpha$ in  Eq.(\ref{eq:Fme}) is given by
\begin{equation}
\alpha  = \frac{1}{\left(1 - g\right)}\frac{3 Q \langle \mu \rangle J_{1}' }{2(J_1+J_2) K v},
\end{equation}
where $Q = -2e$ is the charge of the oxygen ion, $\langle \mu \rangle = 2 \mu_B \langle S \rangle$ is the average magnetic moment on each site, $v$ is the unit cell volume,
and
\begin{equation}
g = \frac{15}{8} \frac{\left(J_{1}'\langle S \rangle\right)^2}{(J_1+J_2) K}
\end{equation}
is the dimensionless spin-lattice coupling constant.

An estimate for the magnetoelectric constant,  $\alpha \sim 10^{-3}$, for the model parameters appropriate for the KITPite structure ($S = 2$, $J_{1} \sim 3$meV, $K \sim 6 \mbox{eV}\cdot\mbox{\AA}^{-2}$, $\frac{J'_{1}}{J_{1}} \sim 3.5 \mbox{\AA}^{-1}$, and $v = 177 \mbox{\AA}^3$) agrees well with the result of ab initio calculations.\cite{DelaneyPRL2009} Furthermore, $\chi_{\rm m} \sim 2\cdot 10^{-4}$, so that $\left( \frac{\alpha}{{\chi}_{\rm m}} \right)^2 \sim 25$. Thus, the electromagnon peak in KITPite should be much stronger than the antiferromagnetic resonance peak, which is also the case for rare earth manganites with a spiral ordering.\cite{ValdesPRL2009}

At $g = 1$, the magnetoelectric constant diverges and so do the dielectric and magnetic susceptibilities:
\begin{equation}
\chi_{e}, \chi_{m} \propto \frac{1}{1-g}.
\end{equation}
Since in our model there are two polar phonons coupled to a magnon with a given polarization (one in the up-triangle and another in the down-triangle), the magnetoelectric constant comes close to its upper bound but does not reach it at $g=1$:
\begin{equation}
\left[\frac{\alpha}{\sqrt{\chi_e \chi_m}}\right]_{g = 1} \approx 0.985.
\end{equation}

Surprisingly, the softening of the $q_0$ magnon mode, which is not coupled to polar lattice distortions, occurs at a lower value $g_{0} < 1$. Since $\omega_0 = 0$, the softening in this case means vanishing velocity of the $q_0$-mode. The velocity vanishes, because away from the $\Gamma$-point in the magnetic Brillouin zone magnons with different symmetry become mixed and the $q_0$-mode is coupled to the electromagnon modes. As the spin-lattice coupling grows and the electromagnon frequency decreases, the lowest-frequency magnon branch is pushed down, which ultimately reduces the velocity of the $q_0$-mode to zero.

\section{Magnetoelastic instabilities}
\label{sec:phasediagram}

\begin{figure}
\includegraphics[width=7cm]{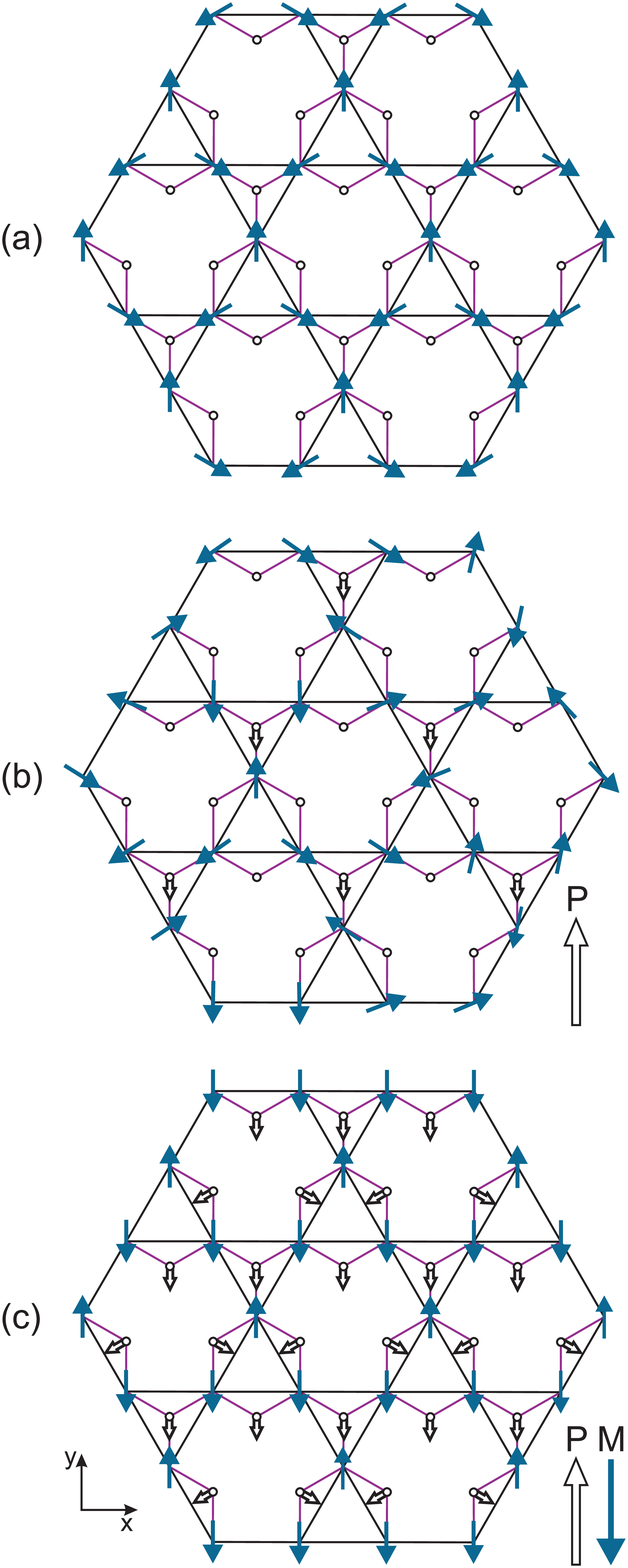}
\caption{(Color online) The minimal-energy spin configurations of the Kagom\'{e} magnet for three different values of the spin-lattice coupling: (a) the 120$^{\circ}$-state with zero wave vector, (b) the incommensurate ferroelectric state, and (c) the collinear multiferroic state. The short solid arrows show the spin directions, while the short empty arrows denote the shifts of the ligand ions in these states. The long solid and empty arrows show the direction of, respectively, the spontaneous magnetization and polarization.} \label{fig:conf}
\end{figure}

Although KITPite, for which $g \sim 0.05$, is far away from the instabilities discussed in the previous section,
it is interesting to study behavior of magnetoelectric materials when the spin-lattice coupling becomes strong, in particular, in view of the dramatic magnetoelectric effects recently observed in multiferroics. As the magnetoelectric constant becomes large close to the transition between magnetoelectric and multiferroic states, it is important to understand possible scenarios of such a transition.

In this section we present the
analytical and numerical study of the phase diagram of the KITPite layer for strong spin-lattice couplings. In particular, we show that none of the continuous transitions involving magnon softening, discussed in Sec.~\ref{sec:softening}, actually takes place, as the strong spin-lattice coupling makes the frustrated Kagom\'{e} magnet unstable towards a first-order magnetoelastic transition that relieves the frustration.
This frustration-driven instability is similar to the one found in spinels, where a collinear ordering of spins appears together with a lattice deformation.\cite{LeePRL2000,TchernyshyovPRL2002,TchernyshyovPRB2002,PencPRL2004} We show that the transformation of a non-collinear magnetoelectric state into a collinear multiferroic state  can involve two transitions and an intermediate phase, which is ferroelectric but not ferromagnetic.

To understand the origin of magnetoelastic instabilities  at strong spin-lattice coupling, we first consider a single up-triangle and integrate out the lattice degrees of freedom, $\mathbf{u}_{i}$ ($i = 1,2,3$). Then the total  energy of the triangle takes the form of an effective spin Hamiltonian with quadratic and bi-quadratic interactions:
\begin{equation}
E_{\triangle} =   \sum_{\langle i,j \rangle}\left[
J_1 \mathbf{S}_i \cdot \mathbf{S}_j
- \frac{\left(J'_{1}\right)^2}{2K}
(\mathbf{S}_i \cdot \mathbf{S}_j)^2
\right].
\end{equation}
The bi-quadratic interactions favor collinear spins and for
${\tilde g} = \frac{15}{8} \frac{\left(J_{1}'S\right)^2}{J_1 K} =
\frac{\left(J_1+J_2\right)}{J_1} g > \frac{5}{6}$, the
lowest-energy spin configuration is a collinear state of the
$\uparrow \uparrow \downarrow$ type (the spins lie in the lattice
plane). Similarly, an effective spin Hamiltonian for a
down-triangle, where the exchange along all bonds is mediated by a
single ligand ion located inside the triangle, has the form,
\begin{eqnarray}
E_{\nabla} &=&   \sum_{\langle i,j \rangle}\left[
J_1 \mathbf{S}_i \cdot \mathbf{S}_j
- \frac{\left(J'_{1}\right)^2}{2K}
(\mathbf{S}_i \cdot \mathbf{S}_j)^2
\right] \nonumber \\ &~& + \frac{\left(J'_{1}\right)^2}{2K}
\sum_{i \neq j \neq k} \left(\mathbf{S}_i \cdot \mathbf{S}_j\right) \left(\mathbf{S}_j \cdot \mathbf{S}_k\right).
\end{eqnarray}
In this case the critical coupling is lower: ${\tilde g} = \frac{15}{32}$. Due to the inequivalence of the up- and down-triangles in the KITPite structure the transition from the $120^\circ$-state shown in Fig.~\ref{fig:conf}a to a fully collinear state shown in Fig.~\ref{fig:conf}c goes in two steps via an intermediate state, where only the spins in the down-triangles are collinear while the spins in the up-triangles are still non-collinear (see Fig.~\ref{fig:conf}b).

\begin{figure}[htbp]
\includegraphics[width=8cm]{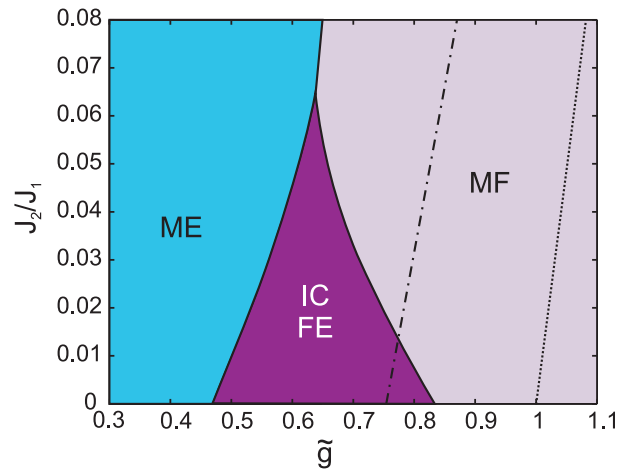}
\caption{(Color online) Plotted is the zero-temperature phase
diagram of the Kagom\'{e} layer for large values of the spin-lattice
coupling ${\tilde g} = \frac{15}{8} \frac{\left(J_{1}'\langle S
\rangle\right)^2}{J_1 K} = \frac{\left(J_1+J_2\right)}{J_1} g$.
For relatively small  $\frac{J_2}{J_1}$ the magnetoelectric (ME)
phase with the 120$^{\circ}$ spin ordering [see
Fig.~\ref{fig:conf}(a)] undergoes a first-order transition into
the incommensurate (IC) spin state, which is ferroelectric (FE)
[see Fig.~\ref{fig:conf}(b)], as the coupling constant ${\tilde
g}$ increases.  Further increase of ${\tilde g}$ results in the
transition into the collinear multiferroic (MF) phase [see
Fig.~\ref{fig:conf}(c)], with the parallel spontaneous
polarization and magnetization, $\mathbf{P} \parallel \mathbf{M}$.
For larger $\frac{J_2}{J_1}$, the ME state undergoes a direct
transition into the fully collinear MF state. Also plotted are the
dot-dash line, where the $q_0$-mode in the ME state softens and
the dotted line, at which the magnetoelectric response of the ME
state diverges (both phenomena do not occur because of the
intervening first-order transitions to the FE and MF states).}
\label{fig:pd}
\end{figure}

The transition to the intermediate state does not fully lift the frustration: the number of combinations of the down-triangles with collinear spins and up-triangles with the $120^{\circ}$-angle between spins grows exponentially with the system size. In our model this degeneracy is removed by the next-nearest-neighbor interactions between spins, which select the state shown in Fig.~\ref{fig:conf}b. The next-nearest-neighbor interactions in the vertical direction induce a small spin canting, as a result of which spins in down-triangles are not strictly collinear and the angles between spins in up-triangles deviate from $120^{\circ}$ by the angle $\varphi \propto \frac{J_2}{J_1}$. Furthermore, the nearest-neighbor interactions in the remaining two directions are frustrated in the commensurate spin state with the wave vector $\mathbf{Q} = 2k\left(1,0\right)+k\left(\frac{1}{2},\frac{\sqrt{3}}{2}\right)$, where $k = \frac{\pi}{3a}$ ($a$ being the lattice constant). This frustration is lifted, when the spin ordering becomes incommensurate with the lattice:
$k = \frac{\pi}{3a} + \delta$, where $\delta \propto \frac{J_2}{J_1}$.

The incommensurate state has zero net magnetization, but its electric polarization is nonzero. For the state shown in Fig.~\ref{fig:conf}(b) the polarization vector is parallel to the $y$ axis. This polarization originates not from the inverse Dzyaloshinskii-Moriya mechanism, which makes incommensurate spiral states in e.g. $R$MnO$_3$ ferroelectric, but from the fact that all bonds connecting parallel spins are parallel to each other, which forces the ligand ions in all down-triangles to shift in the same direction, since
\begin{equation}
\left\{
\begin{array}{rcl}
v_{x} &=& \eta \frac{1}{\sqrt{2}}
\left(\mathbf{S}_3 \cdot \mathbf{S}_5 - \mathbf{S}_3 \cdot \mathbf{S}_4 \right), \\ \\
v_{y} &=& \eta
\frac{1}{\sqrt{6}}
\left(
2\left(\mathbf{S}_4 \cdot \mathbf{S}_5\right)  -
\mathbf{S}_3 \cdot \mathbf{S}_4 - \mathbf{S}_3 \cdot \mathbf{S}_5 \right),
\end{array}
\right.
\end{equation}
where $\eta = \sqrt{\frac{3}{2}}
\frac{J'_{1}}{K}$ and the labeling of spins is the same as in Fig.~\ref{fig:model}. For the state shown in Fig.~\ref{fig:conf}(b), where bonds connecting (nearly) parallel spins are oriented along the $x$ axis, $v_x = 0$, while $v_y$ and hence the polarization $P_y$ is nonzero.

We note that the coupling of exchange interactions to strains, which gives rise to magnetoelastic transitions in frustrated spinels, \cite{LeePRL2000,TchernyshyovPRL2002,TchernyshyovPRB2002}  also favors the ferroelectric state. The absence of inversion symmetry in the KITPite layer allows for the piezoelectric coupling,
\begin{equation}
2 u_{xy} E_x + (u_{xx}-u_{yy}) E_y,
\end{equation}
where $u_{ij}$ is the strain tensor, so that the ligand displacement, $\mathbf{v}$, the electric polarization, $\mathbf{P}$, and the strains are coupled to each other.

The zero-temperature phase diagram of the Kagom\'{e} layer with
${\tilde g}$ and  $\frac{J_2}{J_1}$ along the horizontal and
vertical axes is shown in Fig.~\ref{fig:pd}. For small
$\frac{J_2}{J_1}$, the incommensurate ferroelectric (IC FE) state,
discussed above, intervenes between the magnetoelectric (ME) state
with the 120$^{\circ}$ spin ordering [see Fig.~\ref{fig:conf}(a)]
and the fully collinear multiferroic (MF) state shown in
Fig.~\ref{fig:conf}(c), in which the spontaneous polarization and
magnetization are parallel to each other, $\mathbf{P} \parallel
\mathbf{M}$. As the ratio $\frac{J_2}{J_1}$ grows, the interval of
the coupling constant ${\tilde g}$ where the intermediate state is
stabilized shrinks and above the tricritical point the ME state
undergoes a direct transition into the collinear MF state along
the critical line $\frac{J_2}{J_1} = \frac{5}{3} {\tilde g} - 1$.
Also plotted are the dot-dash line, at which the $q_0$-mode would
soften and the dotted line, at which $\alpha$, $\chi_{\rm e}$, and
$\chi_{\rm m}$ would diverge, if the  ME state would survive at
strong spin-lattice couplings.

\section{Discussion}
\label{sec:discussion}

We showed that the static magnetoelectric response of non-collinear antiferromagnets can be related to hybrid magnon-phonon modes coupled to both electric and magnetic fields. Such magnetoelectric materials are analogs of displacive ferroelectrics the dielectric response of which is governed by optical phonon modes. If spins in an ordered state are collinear, the exchange striction cannot couple an electric field to a single magnon, as the expansion of scalar products of parallel or antiparallel spins begins with terms of second order in $\delta \mathbf{S}$, which give rise to photoexcitation of a  two-magnon continuum (the so-called ``charged magnons"\cite{DamascelliPRL1998}). Electromagnons in collinear magnets can still originate from mechanisms  involving relativistic effects, such as the exchange striction induced by the antisymmetric Dzyaloshinskii-Moriya interaction, which is  proportional to the vector product of two spins. In $3d$ transition metal compounds such couplings are weak compared to the exchange striction driven by the Heisenberg exchange, so that the spectral weight of electromagnons in collinear magnets should be relatively low.

Magnetoelectric materials with collinear spin orders may rather be analogs of `order-disorder' ferroelectrics with  the static magnetoelectric response originating from  thermal spin fluctuations. Cr$_2$O$_3$ seems to be an example of such a material: its magnetoelectric coefficient passes through a maximum below N\'{e}el temperature and then strongly decreases when temperature goes to zero and spin fluctuations become suppressed.\cite{RadoPR1962,YatomPR1969}

We note that the rotationally invariant coupling
Eq.(\ref{eq:Hme1}) may also originate from purely electronic
mechanisms, such as the polarization of electronic orbitals
induced by a magnetic ordering.
\cite{SergienkoPRL2006,PicozziPRL99,BulaevskiiPRB2008,SpaldinJPCM2008,Furukawa2009}
Ab initio calculations suggest that in rare earth manganites the
electronic mechanisms of magnetoelectric coupling are as important
as the exchange striction.\cite{PicozziPRL99} On the other hand, the increase of the spectral weight of the electromagnon peaks in $R$MnO$_3$ below the spiral ordering temperature occurs largely at
the expense of the strength of the optical phonon peak at $\sim 100$cm$^{-1}$,
suggesting the dominant role of the spin-lattice coupling.
\cite{SushkovJPCM2008,ValdesPRL2009} If electronic mechanisms dominate and  an electromagnon gets its spectral weight from
frequencies much higher than those of optical phonons,
Eq.(\ref{eq:KK}) should to be modified in an obvious way, while
Eq.(\ref{eq:emagafmr}), where $\frac{\omega_{\rm mag}}{\omega_{\rm
ph}}$ should be replaced by $0$, is still valid.

We note that the non-collinearity of spins by itself does not guarantee strong magnetoelectric effect and electromagnon peaks -- the crystal structure is equally important. Thus, in the layered Kagom\'{e} antiferromagnet, the iron jarosite KFe$_3$(OH)$_6$(SO$_4$)$_2$,\cite{GroholNatMat2005} which has the spin ordering shown in Fig.~\ref{fig:model}, the ligand ions are located outside of both up- and down-triangles, which cancels the magnetoelectric effect due to the Heisenberg exchange striction. The cancellation also occurs in triangular magnets with the $120^{\circ}$ spin ordering, as they contain three different spin triangles, such that spins in one triangle are rotated by $\pm 120^{\circ}$ with respect to spins in two other triangles\cite{DelaneyPRL2009} (more generally, the linear magnetoelectric effect can only be induced by a spin ordering with zero wave vector). We note, however, that the lattice trimerization in hexagonal manganites\cite{AkenNatMat2004} makes the  three types of spin triangles inequivalent and destroys the cancellation. This can be also seen from the symmetry properties of the A$_{1,2}$ and B$_{1,2}$ phases of hexagonal manganites\cite{FiebigJAP2003} allowing for the magnetoelectric term  $E_x H_y - E_y H_x$ in the A$_1$-phase, which has a toroidal moment, and the term $E_x H_x + E_y H_y$ in the A$_2$-phase, which has a magnetic monopole moment. Whether electromagnons in these phases can be observed, depends on the magnitude of the trimerization and remains to be explored.\cite{LeeNature2008}

\section{Conclusions}
\label{sec:conslusions}

In conclusion, we showed that magnets with non-collinear spin orders resulting in a linear magnetoelectric effect may also show electromagnon peaks in optical absorption spectrum. While electromagnons should be present in many non-collinear magnets, the specific feature of magnetoelectric materials is that some magnon modes can be excited by both electric and magnetic fields, i.e. electromagnons are also antiferromagnetic resonances. We derived a simple relation Eq.(\ref{eq:emagafmr}) between the ratio of the spectral weights of the electromagnon and antiferromagnetic resonance peaks and the ratio of the static  magnetoelectric constant and magnetic susceptibility,
which can be used to estimate the strength of electromagnon peaks on the basis of dc measurements.

To make our consideration more specific, we considered a Kagom\'{e} lattice magnet with the KITPite structure, where the ligand ions are positioned in a way that gives rise to a relatively strong linear magnetoelectric effect.\cite{DelaneyPRL2009} Using the symmetry
analysis we identified the magnon modes that are coupled to both electric and magnetic fields and give rise to the linear
magnetoelectric effect. We showed that the softening of these
modes at a strong spin-lattice coupling results in the divergence
of the magnetoelectric constant as well as of magnetic and dielectric
susceptibilities, signaling the instability of the magnetoelectric
state towards a multiferroic state with spontaneously generated
$\mathbf{P}$ and $\mathbf{M}$. However, the detailed study of the phase diagram of this model revealed that the electromagnon softening does not actually take place, since the first-order transition to
the collinear multiferroic state occurs at a lower value of the spin-lattice coupling. In some region of model parameters a ferroelectric incommensurate-spiral phase intervenes between the magnetoelectric and multiferroic phases. While in known spiral magnets, ferroelectricity is likely induced by the inverse Dzyaloshinskii-Moriya mechanism, in our model it results from the stronger exchange striction mechanism due to the collinearity  of spins in half of the triangles. These magnetoelastic instabilities are typical for frustrated magnets, where non-collinear spin orders usually occur. We also
discussed the possibility to observe electromagnons in known magnetoelectric materials. We hope that our study will stimulate experimental work in this direction.

\begin{acknowledgments}
This work was supported by the Zernike Institute for Advanced Materials and by the Stichting voor Fundamenteel Onderzoek der Materie (FOM).
\end{acknowledgments}

\end{document}